# Fast domain wall motion induced by antiferromagnetic spin dynamics at the angular momentum compensation temperature of ferrimagnets

Kab-Jin Kim[1,2†★], Se Kwon Kim[3†], Takayuki Tono[1], Se-Hyeok Oh[4], Takaya Okuno[1], Woo Seung Ham[1], Yuushou Hirata[1], Sanghoon Kim[1], Gyoungchoon Go[5], Yaroslav Tserkovnyak[3], Arata Tsukamoto[6], Takahiro Moriyama[1], Kyung-Jin Lee[4,5,7★], and Teruo Ono[1★]

[1]*Institute for Chemical Research, Kyoto University, Gokasho, Uji, Kyoto, 611-0011, Japan*

[2] *Department of Physics, Korea Advanced Institute of Science and Technology, Daejeon 34141, Korea*

[3]*Department of Physics and Astronomy, University of California, Los Angeles, California 90095, USA*

[4]*Department of Nano-Semiconductor and Engineering, Korea University, Seoul 02841, Korea*

[5]*Department of Materials Science & Engineering, Korea University, Seoul 02841, South Korea*

[6]*College of Science and Technology, Nihon University, Funabashi, Chiba 274-8501, Japan*

[7]*KU-KIST Graduate School of Converging Science and Technology, Korea University, Seoul 02841, South Korea*

★ Correspondence to: kabjin@scl.kyoto-u.ac.jp, kj_lee@korea.ac.kr, ono@scl.kyoto-u.ac.jp

**Antiferromagnetic spintronics is an emerging research field which aims to utilize antiferromagnets as core elements in spintronic devices[1,2]. A central motivation toward this direction is that antiferromagnetic spin dynamics is expected to be much faster than ferromagnetic counterpart because antiferromagnets have higher resonance frequencies than ferromagnets[3]. Recent theories indeed predicted faster dynamics of antiferromagnetic domain walls (DWs) than ferromagnetic DWs[4-6]. However, experimental investigations of antiferromagnetic spin dynamics have remained unexplored mainly because of the immunity of antiferromagnets to magnetic fields. Furthermore, this immunity makes field-driven antiferromagnetic DW motion impossible despite rich physics of field-driven DW dynamics as proven in ferromagnetic DW studies. Here we show that fast field-driven antiferromagnetic spin dynamics is realized in ferrimagnets at the angular momentum compensation point $T_A$. Using rare-earth–3d-transition metal ferrimagnetic compounds where net magnetic moment is nonzero at $T_A$, the field-driven DW mobility remarkably enhances up to 20 km s$^{-1}$T$^{-1}$. The collective coordinate approach generalized for ferrimagnets[7] and atomistic spin model simulations[6,8] show that this remarkable enhancement is a consequence of antiferromagnetic spin dynamics at $T_A$. Our finding allows us to investigate the physics of antiferromagnetic spin dynamics and highlights the importance of tuning of the angular momentum compensation point of ferrimagnets, which could be a key towards ferrimagnetic spintronics.**

Encoding information using magnetic DW motion is essential for future magnetic memory devices, such as racetrack memories[9,10]. High-speed DW motion is a key prerequisite for making the racetrack feasible. However, velocity breakdown due to the angular precession of DW, referred to as the Walker breakdown[11], generally limits the functional performance in ferromagnet-based DW devices. Recently, it was reported that the DW speed boosts up significantly in antiferromagnets due to the suppression of the angular precession[4-6]. However, the immunity of antiferromagnets to magnetic fields yields notorious difficulties in creating, manipulating, and detecting antiferromagnetic DWs, compared to ferromagnetic ones. One possibility to avoid these difficulties is offered by the synthetic antiferromagnets[12], where the net magnetic moment can be controlled by tuning the thickness of two ferromagnetic layers coupled antiferromagnetically. However, they still suffer from the field-immunity when the net magnetic moment approaches zero, preventing the study of antiferromagnetic DW dynamics. Here we show that magnetic field-controlled antiferromagnetic spin dynamics can be achieved by employing ferrimagnets.

There are a class of ferrimagnets, rare earth (RE)–transition metal (TM) compounds, where the spins of two inequivalent sublattices are coupled antiferromagnetically. Because of different Landé g-factors between RE and TM elements, these ferrimagnets have two special temperatures below the Curie temperature: the magnetisation compensation temperature, $T_M$, at which the two magnetic moments cancel each other, and the angular momentum compensation temperature, $T_A$, at which the net angular momentum vanishes[13-15]. In particular, the existence of $T_A$ in ferrimagnets provides a framework to investigate antiferromagnetic spin dynamics. It is because the time evolution of the state of a magnet is governed by the commutation relation of the angular momentum, not of the magnetic moment. As a result, the nature of the dynamics of the ferrimagnets is expected to change from ferromagnetic to antiferromagnetic as approaching the angular momentum compensation point $T_A$. Furthermore, the net magnetic moment of ferrimagnets is nonzero at $T_A$ and can thus couple to an external magnetic field, opening a new possibility of field-driven antiferromagnetic spin dynamics.

In order to pursue this possibility, we first describe distinguishing features of ferrimagnetic DWs near $T_A$ based on the collective coordinate approach[7]. A ferrimagnetic DW effectively acts as an antiferromagnetic DW around $T_A$. The low-energy dynamics of a DW in quasi one-dimensional magnets is generally described by two collective coordinates, its position $X$ and angle $\Phi$, which capture the translational and spin-rotational degrees of freedom of the DW, respectively. In ferromagnets, they are gyrotropically coupled by the Berry phase that is proportional to the net spin density, and the motion of a DW slows down severely above a certain critical stimulus, engendering the phenomenon of the Walker breakdown[11]. In antiferromagnets, on the other hand, the dynamics of $X$ and $\Phi$ are independent owing to vanishing the net spin density[7]. The DW dynamics in antiferromagnets is thus free from the Walker breakdown and can be fast in a broad range of external driving forces compared to that in ferromagnets.

In the following, we explain a theory for the field-driven DW dynamics in ferrimagnets in high fields, which agrees with our experimental results as discussed later. The DW velocity can be derived as follows by invoking the energy conservation and the gyrotropic coupling between the two collective coordinates $X$ and $\Phi$ (see Supplementary Information for microscopic derivations). For a sufficiently strong external field, the anisotropy energy can be neglected and the time derivatives of $X$ and $\Phi$ can be considered to approach constant values[16], $\dot{X} \to V$ and $\dot{\Phi} \to \Omega$. The energy-dissipation rate caused by these dynamics is given by $P = 2(\alpha_1 s_1 + \alpha_2 s_2) \mathsf{A}(V^2 + \lambda^2 \Omega^2)/\lambda$, where $\mathsf{A}$ is the cross sectional area of the magnet, $\lambda$ is the domain-wall width, $\alpha_1$ and $s_1$ are the Gilbert damping constant and the spin angular momentum density of one sublattice, respectively, and $\alpha_2$ and $s_2$ are for the other sublattice. Invoking the conservation of the total energy, the rate of energy dissipation can be equated to the decreasing rate of the Zeeman energy induced by the translation motion of the domain wall, which yields the equation, $2(\alpha_1 s_1 + \alpha_2 s_2) \mathsf{A}(V^2 + \lambda^2 \Omega^2)/\lambda = 2(M_1 - M_2)H\mathsf{A}V$, where $M_1$ and $M_2$ are the magnetization of the two sublattices, and $H$ is an external field. Note that the net magnetization, $M_1 - M_2$, does not vanish at $T_A$ due to the difference in the Landé g-factors of two sublattice atoms, which is essential to drive a DW with an external field. In addition, when there is finite net spin angular momentum $s_1 \neq s_2$, e.g., away from $T_A$, the angular and linear velocities are related by the gyrotropic coupling whose strength is proportional to the net angular momentum $\propto (s_1 - s_2)$ [17]. Balancing the gyrotropic force on $\Phi$ with the dissipative force yields $(\alpha_1 s_1 + \alpha_2 s_2)\lambda\Omega = (s_1 - s_2)V$. Solving the two aforementioned equations for $V$ and $\Omega$, we obtain

$$V = \frac{\lambda(\alpha_1 s_1 + \alpha_2 s_2)(M_1 - M_2)}{(\alpha_1 s_1 + \alpha_2 s_2)^2 + (s_1 - s_2)^2} H, \quad \Omega = \frac{(s_1 - s_2)(M_1 - M_2)}{(\alpha_1 s_1 + \alpha_2 s_2)^2 + (s_1 - s_2)^2} H \tag{1}$$

As the system approaches the angular momentum compensation point $|s_1 - s_2| \to 0$, the domain wall speed $V$ increases, whereas the precessional frequency $\Omega$ decreases. At $T = T_A$, $X$ and $\Phi$ are completely decoupled and the pure translational dynamics of the DW is obtained, implying that the ferrimagnetic DW effectively acts as antiferromagnetic DW and its motion is driven by a magnetic field at $T = T_A$.

In order to prove the above theoretical prediction, we investigate DW dynamics in ferrimagnetic GdFeCo compounds. Figure 1 shows a schematic illustration of our sample. SiN(5 nm)/Gd$_{23}$Fe$_{67.4}$Co$_{9.6}$ (30 nm)/SiN(5 nm) films are deposited on intrinsic Si substrate by magnetron sputtering. GdFeCo is a well-known RE– TM ferrimagnetic compound, in which RE and TM moments are coupled antiferromagnetically[18]. The relative magnetic moments of RE and TM can be easily controlled by varying the composition or temperature, so that $T_M$ and $T_A$ can be easily designed in RE–TM ferrimagnets. The GdFeCo film is then patterned into microwires with 5 μm width and 65 μm length using electron beam lithography and Ar ion milling. A Hall bar is designed to detect the DW motion via the anomalous Hall effect (AHE) voltage, $V_H$.

We first characterise the magnetic properties of the GdFeCo microstrips. Figure 2a shows the hysteresis loops of GdFeCo microstrips at various temperatures. The AHE resistance, $R_H$ ($R_H = V_H /I$), is measured by sweeping the out-of-plane magnetic field, $B_Z$. Square hysteresis loops are clearly observed, indicating that GdFeCo has a perpendicular magnetic anisotropy. The coercivity field, $B_C$, and the magnitude of the Hall resistance change, $\Delta R_H$ ($\Delta R_H = R_H(+B_Z) - R_H(-B_Z)$), are extracted from the hysteresis loops and summarised in Fig. 2b. $B_C$ increases with increasing temperature, but a sudden drop is observed at $T = 220$ K. A sign change of $R_H$ is observed at the same temperature. This is a typical behaviour of ferrimagnets at the magnetisation compensation temperature $T_M$[19]. As $T$ approaches $T_M$, the net magnetic moment converges to zero, and thus, a larger magnetic field is required to obtain a sufficiently high Zeeman energy to switch the magnetisation. Thus, $B_C$ diverges at $T_M$. The sign change of $R_H$ represents additional evidence of $T_M$. The magneto-transport properties of GdFeCo are known to be dominated by FeCo moments because the 4$f$ shell, which is responsible for the magnetic properties of Gd, is located far below the Fermi energy level[20]. Thus, the sign change of $R_H$ indicates a change in the relative direction of the FeCo moments with respect to the magnetic field, which occurs at $T_M$. At $T < T_M$, the Gd moment dominates over the FeCo moment so that the Gd moment aligns along the magnetic field direction. However, at $T > T_M$, the FeCo moment is dominant and thus aligns along the magnetic field direction. Therefore, Fig. 2b allows us to identify $T_M$ for our GdFeCo sample, which is approximately 220 K.

Although $T_M$ can be easily determined by magnetisation or magneto-transport measurements, it is generally not easy to determine $T_A$ because $T_A$ is not related to the net magnetisation but rather to the angular momentum of the system. For GdFeCo, the net magnetisation $\vec{M}$ and angular momentum $\vec{A}$ are written as follows[13,14,21]. $\vec{M} = \vec{M}_{Gd} + \vec{M}_{FeCo}$ and $\vec{A} = \vec{A}_{Gd} + \vec{A}_{FeCo} = \vec{M}_{Gd}/\gamma_{Gd} + \vec{M}_{FeCo}/\gamma_{FeCo}$, where $\vec{M}_{Gd(FeCo)}$ and $\vec{A}_{Gd(FeCo)}$ are the magnetic moment and angular momentum of the Gd (FeCo) sub-lattices, respectively, and $\gamma_{Gd(FeCo)} = g_{Gd(FeCo)}\mu_B/\hbar$ is the gyromagnetic ratio of Gd (FeCo), where $\mu_B$ is the Bohr magneton and $\hbar$ is the reduced Planck constant. According to the literatures[22–24], $g_{FeCo}$ (~2.2) is slightly larger than $g_{Gd}$ (~2) owing to the spin-orbit coupling of FeCo and zero orbital angular momentum of the half-filled 4$f$ shell of Gd; therefore, $T_A$ is expected to be higher than $T_M$ in GdFeCo.

Based on above consideration, we measure the field-driven DW speed at $T > T_M$ using a real-time DW detection method[25–27]. We first saturate the magnetisation by applying a large negative field (|$B$| > |$B_C$|) and then switch the field to the positive direction. This positive field is a DW driving field, $B_d$, and

should be smaller than $B_C$ ($|B_d| < |B_C|$). Next, we inject a current pulse into the electrode to create a DW by a current-induced Oersted field, as shown in Fig. 1. The created DW propagates along the wire due to the presence of $B_d$, and the DW motion is detected at the Hall bar by monitoring the changes in $V_H$. Here, the changes in $V_H$ are recorded by an oscilloscope such that nanosecond time-resolution can be achieved. The DW speed can be calculated from the arrival time and the travel distance (60 μm) of the DW. The details of the measurement scheme are explained in the Method section.

Figure 3a shows the DW speed as a function of $B_d$ at several temperatures above $T_M$. The DW velocity increases linearly with field for all temperatures. Such a linear behaviour can be described by $v = \mu(B_d - B_0)$. Here, $\mu$ is referred to as the DW mobility and $B_0$ is the correction field, which generally arises from imperfections of the sample or complexities of the internal DW structure[28,29]. Figure 3b shows the DW velocity as a function of temperature for several bias fields. The DW velocity shows a sharp peak as expected near $T = T_A$ based on Eq. (1). The DW mobility $\mu$ is estimated from the linear fit in Fig. 3a (dashed lines) and is plotted as a function of the temperature in Fig. 3c. Starting from $T = 260$ K, which is slightly higher than $T_M$, $\mu$ increases steeply, reaching its maximum at $T = 310$ K, and then decreases with a further increase in temperature. The peak mobility is as high as 20 km·s$^{-1}$·T$^{-1}$ at $T = 310$ K. These experimental results are in agreement with the analytical expression Eq. (1), which predicts a Lorentzian shape of the DW velocity near $T = T_A$ with the width $\Delta \sim \alpha_1 s_1 + \alpha_2 s_2$. Such a consistency between experiment and theory manifests that the ferrimagnetic DW indeed acts as an antiferromagnetic DW at $T = T_A$.

We next perform atomic spin model simulations based on the atomistic Landau-Lifshitz-Gilbert (LLG) equation[5,8] (see Method section for details) to verify the experimental result and theoretical prediction. We employ a set of the reduced magnetic moments around $s_1 - s_2 = 0$ as shown in Table 1. The total number of set is 9 in which the index 5 corresponds to the temperature at $T_A$. We assume that a DW is of Bloch-type with perpendicular magnetic anisotropy along the $z$ axis. Figure 4a shows the DW velocity as a function of the external field $B_Z$, applied along the $z$ axis. Velocities increase linearly with $B_Z$ for $B_Z > 10$ mT. The numerical results (circular symbols) are in excellent agreement with the analytic solution (solid lines) for the DW velocity for high fields in Eq. (1). The inset of Fig. 4a shows the DW velocity in low field regimes ($B_Z < 10$ mT). The Walker breakdown occurs in this regime except for the case of $T_A$. The vertical dashed lines represent the Walker breakdown field. Figure 4b shows the DW velocity as a function of $\delta_s = s_1 - s_2$. At $\delta_s = 0$, the DW velocity is the highest, which agrees with the experimental result and theory. This good agreement also supports that field-driven antiferromagnetic spin dynamics is realized in ferrimagnets at $T_A$.

To date, the angular momentum compensation point and its effect on the magnetisation dynamics have often been overlooked in studies of ferrimagnets. Laser-induced magnetisation switching[18,30,31] and magnetic DW motion[32–33], which have been major research themes of ferrimagnets, have mostly been studied without identifying $T_A$. It has been investigated in the context of magnetic resonance or magnetisation switching by current around $T_A$[14,15,34], but a clear identification of spin dynamics at $T = T_A$ have been remained elusive[15]. On the other hand, our results clearly show that the antiferromagnetic spin dynamics is achievable at $T = T_A$. Moreover, such antiferromagnetic spin dynamics can be controlled by magnetic field due to the finite magnetic moment at $T = T_A$, which opens a way of studying field-driven antiferromagnetic spin dynamics. Furthermore, the fact that field-driven DW speed exhibits a sharp and narrow peak at $T_A$ provides a simple but accurate method to determine $T_A$, which has not been possible. We also achieve a fast DW speed near room temperature, opening a possibility for ultra-high speed device operation at room temperature. We expect that such findings are also advantageous for current-induced DW motion in ferrimagnets. A low threshold current density, more than one order of magnitude smaller than that of ferromagnets, has already been demonstrated in ferrimagnets[32–33]. Therefore, by tuning $T_A$, one could obtain high-speed and low power consumed spintronic devices using ferrimagnets, which could even be superior to ferromagnetic systems. To conclude, our work suggests that revealing and tailoring $T_A$, which has not been paid much attention, is crucial for controlling ferrimagnetic magnetisation dynamics, and therefore could be a key for realising ferrimagnetic spintronics.

**Method**

**Film preparation and device fabrication.** The studied samples are amorphous thin films of $Gd_{23}Fe_{67.4}Co_{9.6}$ of 30 nm thickness which have been deposited by magnetron sputtering. To avoid oxidation of the GdFeCo layer, 5 nm $Si_3N_4$ were used as buffer and capping layers, respectively. The films exhibit an out-of-plane magnetic anisotropy. GdFeCo microstrips with a 100 nm-wide Hall bar structure were fabricated using electron beam lithography and Ar ion milling process. A negative tone electron beam resist (maN-2403) was used for lithography at a fine resolution (~5 nm). For current injection, Ti(5 nm)/Au(100 nm) electrodes were stacked on the wire. To make an Ohmic contact, the $Si_3N_4$ capping layer was removed by weak ion milling before electrode deposition.

**Experimental setup for field-driven domain wall motion.** A pulse generator (Picosecond 10, 300B) was used to generate a current pulse to create the DW. 100mA and 10ns current pulse is used to create the DW. For field-driven DW motion, 1mA dc current (corresponding current density is $7 \times 10^9$ A·m$^{-2}$) was flowed along the wire to generate anomalous Hall voltage, $V_H$. Yokogawa 7651 was used as a current source. The $V_H$ at the Hall cross was recorded by the oscilloscope (Textronix 7354) through the 46 dB differential amplifier. Low temperature probe station was used for measuring the DW motion in a wide range of temperature.

**DW detection technique.** We used a time-of-flight measurement of DW propagation to obtain a DW speed in a flow regime. The procedure for measuring the DW speed is as follows. First, a large out-of-plane magnetic field $B_{sat} = -200$ mT is applied to reset the magnetisation. Next, a drive field $B_d$, in the range of $|B_P| < |B_d| < |B_C|$, is applied in the opposite direction. Here, $B_P$ is the pinning field of DW motion and $B_C$ is the coercive field of the sample. Since the $B_d$ is smaller than the $B_C$, the drive field does not reverse the magnetisation or create DWs. Next, a current pulse (100 mA, 10 ns) is injected by a pulse generator to create a DW next to the contact line through current-induced Oersted field. As soon as the DW is created, the $B_d$ pushes the DW because the $B_d$ is larger than the $B_P$. Then the DW propagates along the wire and passes through the Hall cross region. When the DW passes through the Hall cross, the Hall voltage changes abruptly because the magnetisation state of the Hall cross reverses as a result of the DW passage. This Hall signal change is recorded by the oscilloscope through the 46dB differential amplifier. We refer to this as a 'signal trace'. Since the detected Hall voltage change includes a large background signal, we subtract the background from the 'signal trace' by measuring a 'reference trace'. The reference trace is obtained in the same manner as the signal trace, except that the saturation field direction is reversed ($B_{sat}$ = +200 mT). In this reference trace, no DW is nucleated, so that only the electronic noise can be detected in the oscilloscope in the reference trace. To obtain a sufficiently high signal-to-noise ratio, we averaged the data from 5 repeated measurements.

**Atomic spin model simulation.** We adopt the atomistic model simulation since the ferrimagnet consists of two magnetic components, i.e., RE and TM on an atomic scale. The one-dimensional Hamiltonian of ferrimagnet is described by $\mathcal{H} = A_{sim} \sum_i \boldsymbol{S_i} \cdot \boldsymbol{S_{i+1}} - K_{sim} \sum_i (\boldsymbol{S_i} \cdot \hat{\boldsymbol{z}})^2 + \kappa_{sim} \sum_i (\boldsymbol{S_i} \cdot \hat{\boldsymbol{y}})^2$, where $\boldsymbol{S_i}$ is the normalized magnetic moment at lattice site $i$. The odd number of $i$ represents a site for TM, and the even number of $i$ represents a site for RE. $A_{sim}, K_{sim}, \kappa_{sim}$ denote the exchange, easy-axis anisotropy along the z axis, and hard-axis anisotropy, respectively. We solve the atomistic LLG equation $\frac{\partial \boldsymbol{S_i}}{\partial t} = -\gamma_i \boldsymbol{S_i} \times \boldsymbol{H_{eff,i}} + \alpha_i \boldsymbol{S_i} \times \frac{\partial \boldsymbol{S_i}}{\partial t}$, where $\boldsymbol{H_{eff,i}} = -\frac{1}{M_i} \frac{\partial \mathcal{H}}{\partial \boldsymbol{S_i}}$ is the effective field, $\gamma_i = g_i \mu_B / \hbar$ is the gyromagnetic ratio, and $M_i$ is the magnetic moment for site $i$. We use parameters as $A_{sim} = 7.5 meV, K_{sim} = 0.3 meV, \kappa_{sim} = -0.8 \mu eV$, damping constant $\alpha_{TM} = \alpha_{RE} = 0.004$, the lattice constant is 0.4 nm, and Landé g-factors for each site are $g_{TM} = 2.2$ and $g_{RE} = 2$.

**Acknowledgements**

This work was partly supported by JSPS KAKENHI Grant Numbers 15H05702, 26870300, 26870304, 26103002, 25220604, 2604316 Collaborative Research Program of the Institute for Chemical Research, Kyoto University, and R & D project for ICT Key Technology of MEXT from the Japan Society for the Promotion of Science (JSPS). KJK acknowledges support from the KAIST start-up funding. SKK and YT acknowledge the support from the Army Research Office under Contract No. 911NF-14-1-0016. K.-J.L. acknowledges support from Creative Materials Discovery Program through the National Research Foundation of Korea (NRF-2015M3D1A1070465).

**Author contributions**

K.-J.K., T.M., and T.O. planed the study. A.T. grew and optimised the GdFeCo film. T.T. fabricated the device and performed the experiment with the guide of K.-J.K.. T.Okuno, W.-S.H., Y.H., and S.K. helped the experiment. S.-K.K., K.-J.L., and Y.T. provide theory. S.-H.O., G.G., and K.-J.L. performed the numerical simulation. K.-J.K., S.-K.K., K.-J.L., T.M., and T.O analysed the results. K.-J.K., S.-K.K., K.-J.L., T.M., and T.O. wrote the manuscript.

**Additional Information**

Supplementary Information is available in the online version of the paper. Reprints and permissions information is available at www.nature.com/reprints. Correspondence and requests for materials should be addressed to K.-J.K, K.-J.L. and T. O.

**Competing financial interests**

The authors declare no competing financial interests.

**Figure Legends**

**Figure 1| Schematic illustration of Device structure.** Schematic illustration of GdFeCo microwire. The inset shows schematic illustration of two spin sub-lattices below and above the magnetisation compensation temperature, $T_M$. Blue and red arrows indicate Gd and FeCo moments, respectively.

**Figure 2| Identification of magnetisation compensation temperature $T_M$. a,** Anomalous Hall effect resistance $R_H$ as a function of perpendicular magnetic field $B_Z$ for several temperatures as denoted in the figure. **b,.** Coercive field, $B_C$, and the magnitude of the Hall resistance change, $\Delta R_H$ $(\Delta R_H = R_H(+B_Z) - R_H(-B_Z))$ with respective to the temperature. The region shaded in red indicates the magnetisation compensation temperature $T_M$

**Figure 3| Field-driven domain wall (DW) dynamics across the angular momentum compensation temperature $T_A$. a.** DW speed $v$ as a function of driving field $B_d$ for several temperatures as denoted in the figure. Dashed lines are best fits based on $v = \mu(B_d - B_0)$. **b.** DW speed $v$ as a function of temperature $T$ for several driving fields as denoted in the figure. **c.** DW mobility $\mu$ as a function of temperature $T$. The red and blue shaded regions in **b** and **c** indicate the magnetisation compensation temperature, $T_M$, and angular momentum compensation temperature, $T_A$, respectively.

**Figure 4| Simulation results of ferrimagnetic domain wall (DW) a.** DW speed as a function of the out-of-plane field $B_Z$ for various indices (see Table 1). Symbols are numerical results whereas solid lines are Eq. (1). Inset shows low field regimes, where vertical dotted lines indicate the Walker breakdown fields. **b.** Computed DW speed as a function of of $\delta_s = s_1 - s_2$ at various values of $B_Z$.

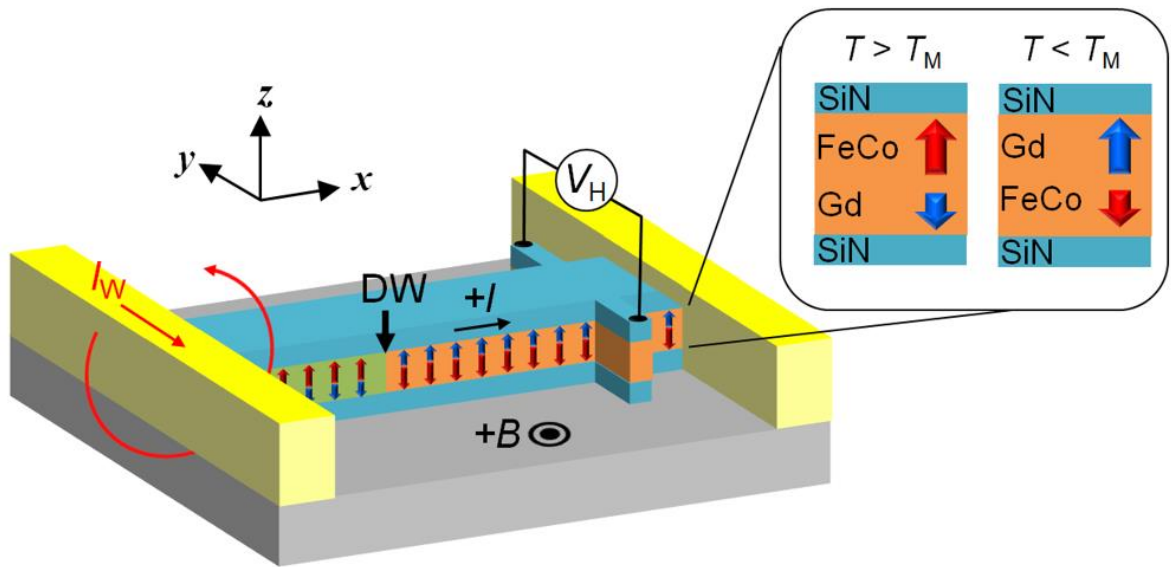

Fig.1

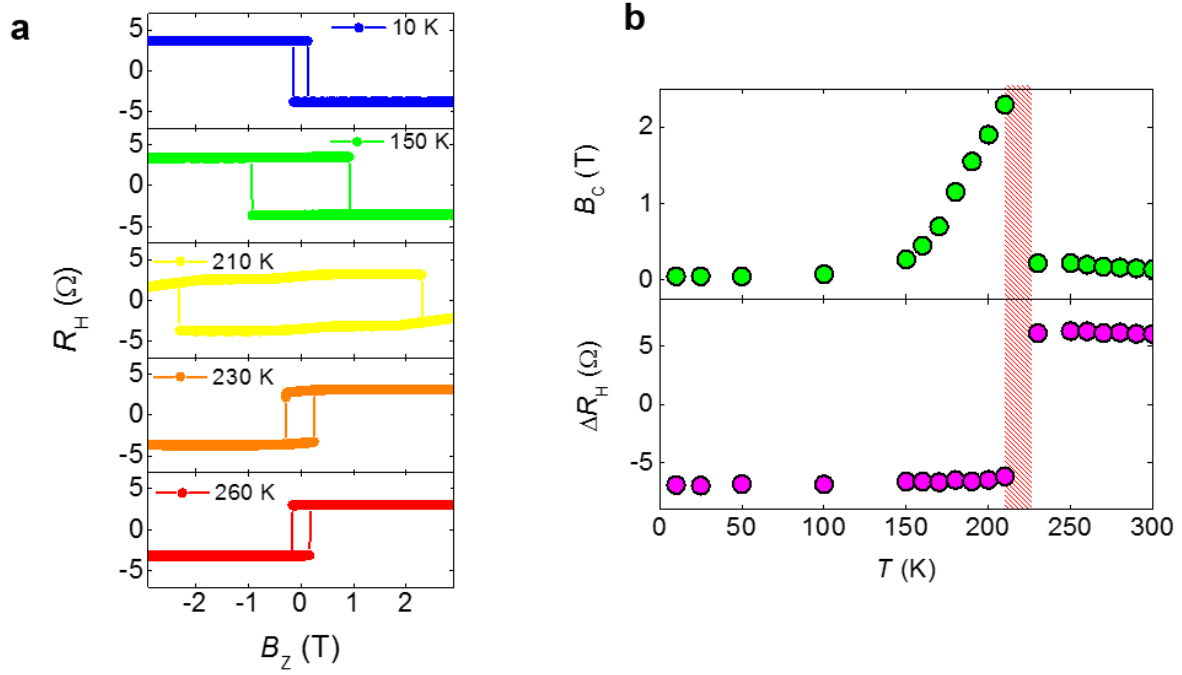

Fig.2

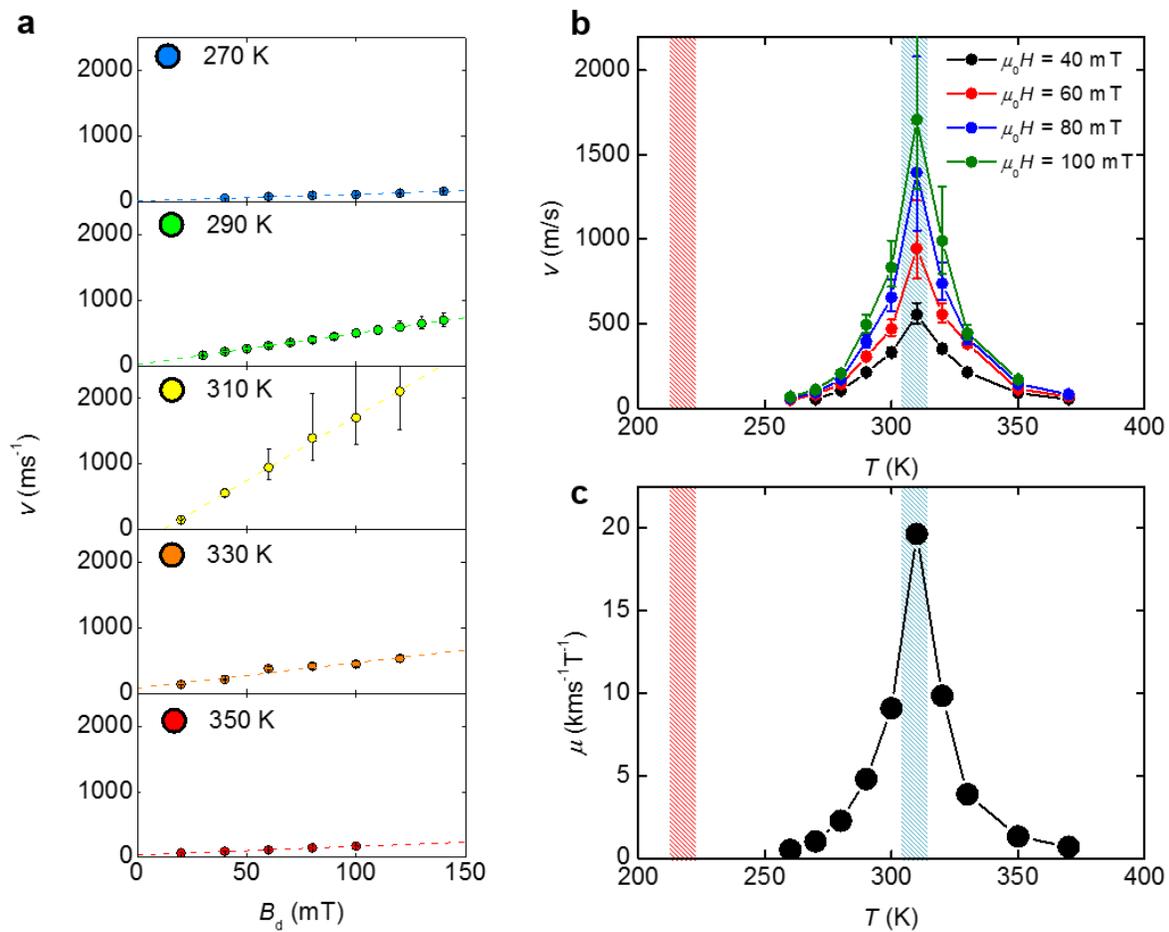

Fig.3

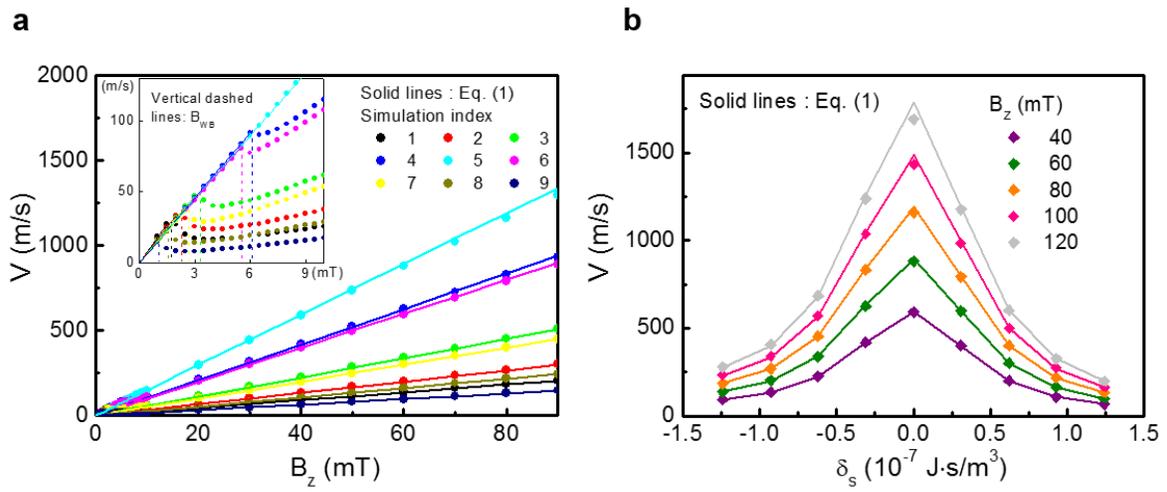

Fig.4

TABLE 1. Parameters used in the numerical simulation.

| Index | 1 | 2 | 3 | 4 | 5 | 6 | 7 | 8 | 9 |
|---|---|---|---|---|---|---|---|---|---|
| $M_{FeCo}(kA/m)$ | 1120 | 1115 | 1110 | 1105 | 1100 | 1095 | 1090 | 1085 | 1080 |
| $M_{Gd}(kA/m)$ | 1040 | 1030 | 1020 | 1010 | 1000 | 990 | 980 | 970 | 960 |
| $\delta_s(10^{-7} J \cdot s/m^3)$ | -1.24 | -0.93 | -0.62 | -0.31 | 0 | 0.31 | 0.62 | 0.93 | 1.24 |